\documentclass[aps,prl,preprint,groupedaddress]{revtex4-1}
\usepackage{bm}
\usepackage{verbatim}   

\usepackage{nopageno}
\usepackage{txfonts}
\usepackage[usenames]{color}

\usepackage{amsfonts,amsbsy,latexsym,amssymb,amscd,amstext}
\usepackage{graphics}
\usepackage{epsfig}
\usepackage{graphicx}
\usepackage{color}      
\usepackage{subfigure}  
\usepackage{caption}

\newcommand{\be}{\begin{equation}}
\newcommand{\ee}{\end{equation}}
\newcommand{\ba}{\begin{array}}
\newcommand{\ea}{\end{array}}
\newcommand{\baa}{\begin{array}}
\newcommand{\eaa}{\end{array}}
\newcommand{\bea}{\begin{eqnarray}}
\newcommand{\eea}{\end{eqnarray}}

\newcommand{\II}{\mathbf{I}}
\newcommand{\Psibar}{\overline{\Psi}}

\begin{document}

\title{Large N meson masses from a matrix model}

\author{  Antonio Gonz\'alez-Arroyo
$^{a,b}$ and Masanori Okawa$^{c,d}$} 
\affiliation{  $^a$ Instituto de F\'{\i}sica Te\'orica UAM/CSIC, \\
       Universidad Aut\'onoma de Madrid, E-28049--Madrid, Spain \\
  $^b$ Departamento de F\'{\i}sica Te\'orica, C-15,  \\
       Universidad Aut\'onoma de Madrid, E-28049--Madrid, Spain \\
  $^c$ Graduate School of Science, Hiroshima University,\\
  Higashi-Hiroshima, Hiroshima 739-8526, Japan \\
  $^d$ Core of Research for the Energetic Universe, Hiroshima University,\\
  Higashi-Hiroshima, Hiroshima 739-8526, Japan 
}

\email{antonio.gonzalez-arroyo@uam.es, okawa@sci.hiroshima-u.ac.jp}

\begin{abstract}
We explain how to compute meson masses in the large $N$ limit using
the twisted Eguchi-Kawai model. A very simple  formula 
is derived,  and we show how it leads in  a fast and efficient way to
results which are in fairly good agreement with other determinations. 
The method is easily extensible to  reduced models with dynamical
fermions based on the twisted reduction idea. 
\end{abstract}



\preprint{IFT-UAM/CSIC-15-106;\hspace*{3mm} \\
FTUAM-15-30;\hspace*{3mm}  \\HUPD-1506}

\date{\today}


\maketitle


 
{\vskip 1cm}
Determining  the properties of gauge theories in the limit of infinite
number of colours is interesting by itself, and not only as a first
term in a $1/N$ expansion. It is a testing ground
where different methodologies can be applied and put in contact and 
contrast. Furthermore, the theory simplifies in this limit and 
certain  properties are simpler to analyze. One example is the role
played by quarks transforming in fundamental representation of the 
group. At large $N$ quark loops are suppressed, so that quark lines only appear as
sources in a pure Yang-Mills theory. Hence, one can consider
observables like Wilson loops in the fundamental representation, and 
they will satisfy an area law relation. In addition, one can also
consider meson states. One expects to have an infinite spectrum of 
stable mesons in this limit.  The situation is perfectly
exemplified by the two dimensional case which was solved by `t
Hooft~\cite{thooftmodel}. In four  dimensions  the result is not known
exactly although there are models and approximations that predict an 
specific result~\cite{evans,pelaez}. It is tempting to use lattice gauge
theories to calculate this spectra. The standard methodology consists
on extracting the masses from correlation functions of quark bilinears in
euclidean space. Since one uses Monte Carlo methods on computers, one
has to work at finite $N$ and then extrapolate the results to infinite
$N$. Furthermore, at small  $N$ quark loops are not suppressed, and one
should in principle work with dynamical quarks, making the computation
highly demanding. Of course, one might also compute the spectra in the
quenched approximation and hope that the approximation becomes
increasingly  accurate as $N$ grows. Although, this approach is perfectly
justified, one should be careful when considering the chiral limit,
since the order of the limits ($m_q \longrightarrow 0$ and
$N\longrightarrow \infty$) might matter, as chiral perturbation
theory is altered in the quenched approximation~\cite{bernard,sharpe}.
Recently, there have been results on the large $N$ meson spectrum 
following this philosophy~\cite{debbio,bali,degrand,lucini}.

In this letter we want to use an alternative approach based on the idea 
of volume independence. If the volume can be kept very small in the
calculation, one can explore much larger values of $N$ with similar 
computer resources. Here in particular we will be using the Twisted
Eguchi-Kawai model (TEK), which is 4-matrix model  which in the large
$N$ limit should be equivalent to the infinite volume large $N$
theory~\cite{TEK1,TEK2,TEK3}. Although the model was proposed long
time ago, a more recent addition constrains the way in which the
chromo-electric and chromo-magnetic discrete flux, characteristic of 
twisted boundary conditions, should be scaled with $N$. With this 
restriction in mind these authors and other collaborators have tested 
the volume reduction hypothesis directly both on the
lattice~\cite{testing} and in the continuum~\cite{string-tension,coupling}. 
Being able to compute the large $N$ meson spectrum within this model 
and comparing it with other determinations is then an interesting  
challenge for the TEK model. Furthermore, the way in which one can do
so, does not look obvious at all. Since the TEK model in obtained 
by reducing the lattice to a single point, one might wonder how can one 
compute correlations of bilinear operators at different points in order 
to compute the spectrum. In addition, readers familiar with the meaning 
of twisted  boundary conditions know that these boundary conditions are
singular for fields in the fundamental representation. Hence, they  might
wonder how one can deal with  quark operators in this context. The
purpose of this letter is precisely to explain these points and produce a
formula which enables one to obtain the meson spectrum from this
matrix model. 

For a state-of-the-art  analysis of the large $N$ spectrum one should
use variational methods with several bilinear quark operators with the 
same quantum numbers. This allows a clear separation of the 
states with the same quantum numbers and a more precise determination
of the ground states and the first few excited ones. This is a fairly
computationally demanding procedure. Hence, in this work we will be
satisfied with explaining the method and presenting some results to 
illustrate whether they are both feasible and reasonable. 
Indeed our formula has a larger range of validity than the
4-dimensional Yang-Mills theory. It can be applied in other dimensions 
and also in theories with dynamical quarks in the adjoint
representation of the group. An extension to the Veneziano limit 
seems also at hand, at least for the case in which the number of
flavours $N_f$ is a multiple of the number of colours $N$.

Here, it is worth mentioning a calculation of some mesonic states
which is the closest in spirit to the one presented
here~\cite{narayanan-neuberger,hietanen}. In these works the authors do not employ 
twisted boundary conditions, but use the idea of partial reduction
introduced previously by some of them~\cite{narayanan-neuberger2} together 
with the so-called {\em quenched momentum prescription}.

The pure gauge theory at large $N$ possesses the volume independence 
property provided the finite volume model is equipped with appropriate 
twisted boundary conditions. This can be taken to the extreme and we
are then led to  the Twisted Eguchi-Kawai model, which is a matrix model where the
dynamical degrees of freedom are given by $d=4$ unitary matrices $U_\mu$. 
In this work, we will restrict to the case of the symmetric twist~\cite{TEK2,TEK3}, 
so that $N$  should be the square of an integer.
The statistical distribution is determined by the Boltzman factor associated to the 
action 
\be
\label{action}
S_{\rm TEK}=-bN \sum_{\mu \ne \nu =0}^{d-1} z_{\nu\mu}{\rm Tr} \left[ U_\mu
  U_\nu U_\mu^\dagger U_\nu^\dagger \right]
\ee
where 
\be
z_{\nu\mu} = \exp \left( k {2\pi i \over \sqrt{N}} \right), \ \ \ z_{\mu\nu}=z_{\nu\mu}^*, \ \ \ \mu>\nu.
\ee
$k$ and $\sqrt{N}$ are co-prime and should satisfy certain
constraints so that the Z($N$) symmetry of the TEK model is not spontaneously broken~\cite{TEK3}.

In  ordinary lattice gauge theory at infinite volume the main observables are the
expectation values of Wilson loops. Given a closed path on the lattice 
${\cal L}$, we can construct a corresponding unitary matrix $U({\cal L})$
by multiplying in  an ordered way the link matrices that define the
path. The Wilson loop expectation values are then given by 
$W({\cal L})= \langle \mathrm{Tr}(U({\cal L})\rangle$. Volume
independence  implies that at large $N$ these values are reproduced by 
the following expectation values in the TEK model: 
\be
\tilde{W}({\cal L})=z({\cal S})\, \langle \mathrm{Tr}(U({\cal L})) \rangle
\ee
In this case ${\cal L}$ represents just an ordered
sequence of direction indices (with orientation) in the same order as for the closed loop 
${\cal L}$ of the ordinary theory. The factor $z({\cal S})$  is given
by the product of the $z_{\mu \nu}$ factors for all plaquettes paving
a surface ${\cal S}$ bounded by the loop ${\cal L}$. This factor does not depend 
on the chosen surface. 

In the weak coupling limit, as $b$ goes to infinity, the expectation
values of  single trace wilson loops of the standard SU($N$) theory tend
towards the trace of the unit  matrix. In the reduced model the path
integral in that limit is dominated by the minima of the action, achieved for 
$U_\mu=\Gamma_\mu$. The matrices $\Gamma_\mu$, called twist-eaters,  
are specific SU($N$) matrices satisfying:
\be
\label{TCR}
\Gamma_\mu \Gamma_\nu = z_{\mu \nu} \Gamma_\nu
\Gamma_\mu 
\ee
These matrices are unique up to  global gauge 
transformations and multiplication by elements of the center. 
To be specific for the rest of the letter we will choose these 
twist-eaters to satisfy $(\Gamma_\mu)^{\sqrt{N}}=\II$.
It is clear that the equivalence of the TEK model in that limit is 
obtained since
\be
\label{zSformula}
\mathrm{Tr}(\Gamma({\cal L}))= N\,  z^*({\cal S})
\ee
where the symbol $\Gamma({\cal L})$ is obtained by replacing $U_\mu$ by 
$\Gamma_\mu$ in the expression of $U({\cal L})$. 
The reader is invited to consult the literature on the TEK model for a proof of the
previous statements.  The equality of expectation values  of rectangular 
Wilson loops for the infinite volume theory and the corresponding 
reduced model observables has been 
verified with great accuracy for a large range of values of the
coupling $b$~\cite{testing}.

We  will now consider, in addition to the gauge fields,
lattice fermions in the fundamental representation. 
In this work we will be using Wilson fermions. This does not seem
essential, so one could easily obtain similar results using staggered 
or overlap fermions. These fermions are not dynamical, but simply act
as sources for the gauge fields. This quenched approximation is 
justified since their dynamical role is suppressed in the large $N$ limit.  

We will now concentrate our efforts in computing correlations of
meson operators. We will focus upon  ultralocal bilinear quark operators
$O_A(x) \equiv \Psibar(x) A \Psi(x)$ with $A$ an element of the
Clifford algebra of spinors. In computing the  correlation function 
of two such operators in ordinary SU($N$) gauge theory, one can
integrate out the contribution of the fermions and express the
result as an expectation value over the gauge fields of a product 
of factors involving the inverse of the Dirac operator and its
determinant. The determinant contribution is suppressed, as mentioned
earlier,  so we are  left with an expression involving the product
of two inverse Dirac operators (quark propagators). Usually, there are two types of
contributions, one of which only contributes to singlet operators. In the
large $N$ limit, factorization also eliminates this contribution to the
meson spectrum, so that the situation simplifies considerably. The
role of flavour is quite limited and appears through the possibility of 
assigning different quark masses to different flavours. Here we will
ignore this distinction since it does not affect our arguments and
we will consider a single flavour of quarks. 

A rather fast way to justify the possibility of obtaining  meson
spectra from the reduced model is to use the hopping parameter 
expansion. This follows by expanding  the quark propagator as a power
series in $\kappa$. The meson two point correlation function takes the
form 
\be
\langle O_A(0) O_B(x) \rangle = \sum_{\gamma\circ\gamma'} R_{A B}\, 
W(\gamma\circ\gamma') 
\ee
where the sum extends over all lattice paths $\gamma$ going from the
origin to the point $x$ followed by  return paths $\gamma'$. The quantity
$R_{A B}$ is the trace of a $4\times 4$ matrix in spinor space, which
contains a factor $\kappa^{|\gamma\circ\gamma'|}$, where $\kappa$ is the 
hopping parameter and $|\gamma\circ\gamma'|$ the length of the
closed path $\gamma\circ\gamma'$.
As explained earlier, the quantity $ W(\gamma\circ\gamma')$ is the  
expectation value of the corresponding Wilson loop in the pure
Yang-Mills theory. Thus, if the reduced model  is capable of
reproducing the expectation value of the Wilson loops, it should also
permit the computation of meson correlators. However, we recall that 
one should replace the Wilson loop expectation values by $\tilde{W}$.
Notice that   to obtain this quantity, it is not enough to identify all links in the
$\mu$ direction with the matrix $U_\mu$, one should also introduce 
the factor $z({\cal S})$. Using the complex conjugate of
Eq. \ref{zSformula}, we see this factor can be obtained by computing the same Wilson
loop but replacing the link matrices by $\Gamma_\mu^*$, the complex
conjugates of the twist-eaters.  Summing up, we see that equivalence
implies that we can replace  the expectation values of the Wilson loops 
of the infinite volume theory  by $1/N$ times the expectation
value of the  Wilson loop obtained by replacing $U_\mu(x)$ by 
$U_\mu\otimes \Gamma_\mu^*$. Resumming the hopping parameter expansion
we would obtain the same formula for the correlation function
that we will be presenting below. 

After this warm up,  let us approach the problem from a more standard
way. The two problems that we signalled in the introductory paragraphs, the
oddity of looking at correlations at different points for a 1-point
box, and the conflict of twisted boundary conditions with 
quarks in the fundamental representation, can be circumvented with the
same idea: quarks should be allowed to propagate in a bigger (even
infinite) lattice. Thus, reduction  is only applied for the gauge
field. The situation resembles what happens in solid state physics
where electrons propagate in a periodic potential. This time, however, 
gauge fields are not quite periodic, but only periodic up to a twist. 
Let us consider the gauge field at one lattice point $n$, $V_\mu(n)$. 
One can translate this gauge field in the $\nu$ direction by applying 
some twist-matrices. We will choose these twist-matrices to be the
twist-eaters given earlier $\Gamma_\nu$. Then we have
\be
V_\mu(n+\hat{\nu})=\Gamma_\nu V_\mu(n) \Gamma_\nu^\dagger
\ee
The non-triviality of the twisted boundary conditions follows from the non-commutativity of
the twist matrices, which satisfy Eq.~\ref{TCR}. 
Although several choices of the twist tensor $z_{\mu \nu}$ are possible, 
as explained earlier, here we will restrict to the case of the symmetric
twist, so that we should take $N$ to be the square of an integer. 

Thus, if we start with the gauge field at one particular point $n=0$ 
$V_\mu\equiv V_\mu(0)$, we can apply translations in different
directions to construct the gauge link at another point $n$ as follows:
\be
V_\mu(n)=\Gamma(n) V_\mu \Gamma^\dagger(n)
\ee
where the $\Gamma(n)$ can be chosen to be 
\be
\Gamma(n)= \Gamma_0^{n_0} \Gamma_1^{n_1}\Gamma_2^{n_2}\Gamma_3^{n_3}
\ee
The matrices $\Gamma(n)$ are all linearly independent for all $n$,
except when $n_\mu$ is proportional to $\sqrt{N}$ in all directions.
In that case these matrices are equal to the identity, implying 
strict periodicity of the gauge field in a box of size $(\sqrt{N})^4$. 
Obviously, in the large $N$ limit the size grows
indefinitely, but at finite $N$ one expects to find corrections taking
the form of finite volume corrections. For the reduced model to be 
a good approximation to the infinite volume theory one needs that all 
correlations lengths are smaller than $\sqrt{N}$.

The links appearing in the reduced model action Eq.~\ref{action} are not the 
$V_\mu$, but rather $U_\mu=V_\mu \Gamma_\mu$. Notice that,  once we
make this change of variables, the plaquette of the gauge
field becomes
\be
\mathrm{Tr}\left(V_\mu(n)V_\nu(n+\mu)V^\dagger_\mu(n+\nu)V^\dagger_\nu(n)\right)=
z_{\nu \mu} \mathrm{Tr}\left(U_\mu U_\nu U_\mu^\dagger
U_\nu^\dagger\right)
\ee
which is the  standard form appearing in  the TEK action.

According to our general idea, we can add quark fields in the
fundamental representation $\Psi(n)$ and
couple them to the $V_\mu(n)$ gauge fields. The choice of the
fundamental representation is not essential and the whole thing could
be repeated for other representations, but  for definiteness we will stick
to the fundamental one in this letter. In principle, these quarks fields 
could live in an infinite lattice, but given that gauge fields live in
an effective finite box, we can take the quarks to live also in this
box. Indeed, for the sake of getting better correlations we will make
the box longer in the time direction. Thus, quark fields are defined as 
periodic  in a box  of size $(\sqrt{N})^3 \times (l_0 \sqrt{N})$ where 
$l_0$ is an integer.  This eliminates the  conflict of boundary
conditions in this or any other representation. In fact, this is not the 
smallest box for which  there is consistency with the twisted boundary 
conditions of the gauge field,  but it is a more  symmetric choice, which we
will then adopt. For this  choice the number of
quark degrees of freedom is $4\times N\times l_0 N^2$, coming from
multiplying spinor indices, colour indices and spatial indices.

 Now we can write down the quark fermionic action for Wilson fermions in
 the standard way: 
 \be
 S_f= \sum_n \left[\bar{\Psi}(n)\Psi(n)- \kappa\sum_{\mu} \left(
 \bar{\Psi}(n) (r-\gamma_\mu) V_\mu(n)
 \Psi(n+\hat{\mu}) +
\bar{\Psi}(n)  (r+\gamma_\mu) V^\dagger_\mu(n-\hat{\mu})
 \Psi(n-\hat{\mu}) \right) \right]
 \ee
 The simplification of the reduced model is hidden in
 the form of the link variables $V_\mu(n)=\Gamma(n) U_\mu
 \Gamma_\mu^\dagger \Gamma^\dagger(n)$, which are just functions of 
 the 4 space-time-independent link matrices $U_\mu$.

Let us now make the following change of variables in the fermion
fields:
\be
\Psi(n)=\Gamma(n) \chi(n)
\ee
The new fermion fields are still periodic in space-time with
the same periodicity. In terms of the redefined quark fields the 
Wilson-Dirac operator takes
the form
\bea
D_{n m}=  \delta(n,m) \mathbf{I}- \kappa\sum_{\mu} (
(r-\gamma_\mu)
U_\mu \Gamma^\dagger_\mu \Gamma^\dagger(n)
\Gamma(n+\hat{\mu})\delta(n+\hat{\mu},m)
+\\
 (r+\gamma_\mu) \Gamma^\dagger(n) \Gamma(n-\hat{\mu})
\Gamma_\mu U^\dagger_\mu  \delta(n-\hat{\mu},m)) 
\eea
One can easily realize that the combinations of $\Gamma(n)$ matrices
satisfy
\be
\Gamma^\dagger_\mu \Gamma^\dagger(n) \Gamma(n+\hat{\mu})= e^{i
\omega_\mu(n)} \mathbf{I} 
\ee
where  $\omega_\mu(n)$ is given by
\be
\omega_\mu(n)=\frac{2 \pi k}{\sqrt{N}}\sum_{\nu>\mu}  n_\nu 
\ee

The Dirac operator acts on spinorial indices, colour indices and
space-time indices. We might make this explicit by writing the operator
in terms of tensor products of matrices acting on these spaces:
\be
D= \mathbf{I}\otimes \mathbf{I} \otimes \mathbf{I} - \kappa \sum_\mu 
\left( (r\mathbf{I}-\gamma_\mu)\otimes U_\mu \otimes \tilde{\Gamma}_\mu
+ (r\mathbf{I}+\gamma_\mu)\otimes U^\dagger_\mu \otimes
\tilde{\Gamma}^\dagger_\mu \right)
\ee
where we have introduced the matrices $\tilde{\Gamma}_\mu$ which act on 
the space-time degrees of freedom only. Its matrix elements are given by 
\be
\label{tildegamma}
\tilde{\Gamma}_\mu(n,m)= e^{i \omega_\mu(n)} \delta(n+\hat{\mu},m)
\ee
It is interesting to notice that the matrices satisfy the following 
relations:
\be
\label{comrel}
\tilde{\Gamma}_\mu \tilde{\Gamma}_\nu = z_{\mu \nu}^*
\tilde{\Gamma}_\nu \tilde{\Gamma}_\mu 
\ee
These are the same relations  satisfied by the complex conjugate 
of the twist-eaters corresponding to the gauge field. The difference
is that these are $V\times V$ matrices, where $V=l_0 N^2$ is the lattice volume,
and the ordinary twist-eaters  are $N\times N$ matrices. This can be
further simplified if we separate the time dependence from
the purely spatial one. This is easy given our choice of $\Gamma(n)$.
Indeed, $\tilde{\Gamma}_i$ are diagonal in the time coordinate $n_0$
and independent on $n_0$. On the other hand, $\tilde{\Gamma}_0$ becomes
\be
\tilde{\Gamma}_0(n,m)  \equiv \delta(m_0, n_0+1) 
\tilde{\Gamma}'_0(\vec{n},\vec{m})= \prod_i (\delta(n_i, m_i))\,  e^{2
\pi k
(n_1+n_2+n_3)/\sqrt{N}}\, \delta(m_0, n_0+1)
\ee
Thus, we might simply consider that  $\tilde{\Gamma}'_0(\vec{n},\vec{m})$
are the components of  $N^{3/2}\times N^{3/2}$ matrix acting  only on  the spatial 
indices. For the remaining directions the matrices can be directly
reduced to the same space $\tilde{\Gamma}_i \longrightarrow \tilde{\Gamma}_i'$. The
$\tilde{\Gamma}'_\mu$ matrices  satisfy the same relations (Eq.~\ref{comrel}).

Using general results~\cite{pegniscola}  on the representations of the twisted algebra 
(Eq.~\ref{comrel}), we can conclude that there is a change of basis 
in the spatial coordinates such that 
\be
\tilde{\Gamma}'_\mu =\Omega (\Gamma_\mu^*\otimes D_\mu) \Omega^\dagger 
\ee
where $\Omega$ is  the  unitary matrix which implements the change of basis 
and the $D_\mu$ are $\sqrt{N}\times \sqrt{N}$ unitary diagonal matrices. 
Furthermore, it is easy to see that $(D_\mu)^{\sqrt{N}}=\mathbf{I}$, so
that all the diagonal elements are given by $\sqrt{N}$ roots of unity. 

Of course, the matrices $\Omega$ and $D_\mu$ can be explicitly derived
once a choice is taken for $\Gamma_\mu^*$. We will leave this as an
exercise for the reader. As we will see, this will not play 
a role for the sake of this letter since  we will  only be 
interested in meson correlation functions  of local operators at zero 
spatial momentum. 

The simple temporal structure of our Dirac matrix 
allows us to compute its inverse (the quark propagator) between two 
times as follows:
\be
P(m_0,n_0) = \frac{1}{l_0\sqrt{N}} \Omega \sum_{p_0} e^{i p_0
(m_0-n_0)}\left[\mathbf{I}-\kappa \sum_\mu
\left( (r-\gamma_\mu) U_\mu
\Gamma_\mu^* D'_\mu + (r+\gamma_\mu) U_\mu^\dagger \Gamma_\mu^t
D_\mu^{\prime *}\right)\right]^{-1}\Omega^\dagger
\ee
where $D'_i=D_i$ and $D'_0=D_0 e^{i p_0}$. The temporal
momentum takes values  $p_0=2 \pi n/(l_0 \sqrt{N})$, with integer $n$.

Now the meson correlator is 
\be
C_{A B}(n_0)= \frac{1}{\sqrt{N}^3} \mathrm{Tr}(A P(0,n_0) B P(n_0,0)) =
\frac{1}{l_0^2 N^{5/2}} \sum_{p_0} \sum_{q_0} e^{-i q_0 n_0} \mathrm{Tr}(A X(p_0+q_0)
B X(p_0))
\ee
where 
\be
X(p_0)=\left[\mathbf{I}-\kappa \sum_\mu \left( (r-\gamma_\mu) U_\mu
\Gamma_\mu^* D'_\mu + (r+\gamma_\mu) U_\mu^\dagger \Gamma_\mu^t
D_\mu^{\prime *}\right) \right]^{-1}
\ee
where we recall that $D'_0=D_0 e^{i p_0}$. In the formula for the
correlator we are projecting both operators over zero-spatial momentum. 
This is implicit since the trace affects not only spinor and color
indices, but also spatial indices. This projection also allow us to
eliminate the dependence in $D_\mu$. As mentioned earlier, these
matrices are diagonal and commute with the whole propagator.
Furthermore, the elements of the diagonal are just $\sqrt{N}$ roots 
of unity $z_\mu$. Since $z_\mu \Gamma_\mu^* = U \Gamma_\mu^* U^{-1}$,
the effect of $z_\mu$ disappears when taking the trace. Hence, we can
safely eliminate $D_\mu$ matrices multiplying the expression by
$\sqrt{N}$. A similar conclusion applies for the factor $e^{ip_0}$ 
when $l_0=1$.
For $l_0>1$ it would only be
true if the $Z(N)$ symmetry remains unbroken (not just $Z(\sqrt{N})$).
With these simplifications we end up with our final formula
\be
\label{formA}
C_{A B}(n_0)=\frac{1}{l_0 N^{3/2}} \sum_{q_0} e^{-i q_0 n_0} \mathrm{Tr}(A
\bar{X}(q_0) B \bar{X}(0))
\ee
where 
\be
\label{formB}
\bar{X}(p_0)=\left[\mathbf{I}-\kappa \sum_\mu \left( (r-\gamma_\mu)
e^{i p_0 \delta_{\mu 0}}U_\mu
\Gamma_\mu^*  + (r+\gamma_\mu) e^{-i p_0 \delta_{\mu 0}} U_\mu^\dagger
\Gamma_\mu^t\right)
\right]^{-1}
\ee
For the $l_0>1$ case and broken $Z(N)$ symmetry one needs to average over 
the $p_0=2 \pi s/(l_0 \sqrt{N})$ where $s$ is an integer modulo $l_0$.

A state of the art calculation of the spectrum should involve a
variational method employing several operators with the same quantum
numbers plus the use of smearing techniques to increase the projection
of these operators onto the lowest lying states in the spectrum.
Although such a program is feasible and is presently being considered,
it would certainly involve a long time, many computer resources and a
bigger research team.  At this stage what we want is to make a
exploratory study which is designed to test the performance of the
method, estimate the number of necessary resources and also verify 
whether the  preliminary results match with other determinations. 
In the following, we will  present the results of this  study. 
More details about the procedure and additional  results are given
in our contribution to the proceedings of the 2015 Lattice Conference~\cite{okawaproc}.

We will concentrate only on ultralocal quark bilinears of the form 
$O_A(x)=\Psibar(x)A\Psi(x)$, where $A$ is one element of the Clifford
algebra. With our limited statistics only the pseudoscalar and 
vector channels have sufficiently small errors ($A=\gamma_5,\gamma_i$).
Meson masses would be extracted from Eqs.~\ref{formA}, \ref{formB}. 
We take $N$=289 corresponding to an effective box of size $17^4$.  
We choose $k$=5, which ensures that the Z($N$) symmetry of the theory is not
spontaneously broken~\cite{TEK3}.
Furthermore,  to have a larger extent of correlation functions we doubled the size 
of the fermionic volume, by choosing $l_0=2$. 

Gauge configurations of the TEK model are generated by a recently proposed
over-relaxation Monte Carlo method~\cite{OR}.  Configurations are separated by 1000 sweeps and
meson propagators are averaged over 800 configurations for each
parameter set ($b$ and $\kappa$). The cost in CPU time on a single node
of SR16000 supercomputer for the computation of correlation functions ranges from 130
to 800 hours depending on the value of $\kappa$. Our whole data set
was generated in 7000 node$\times$hours, which can be done within one month using 10 nodes of SR16000.  To calculate the trace appearing in
Eq.~\ref{formA}, we use the $Z(4)$ random source method.
More details on the simulations are given in ref.~\cite{okawaproc}.

We will now present our results. We have studied the system at two
values of $b=0.36$ and $0.37$ and various values of $\kappa$: 
0.153, 0.154, 0.155,0.156, 0.157, 0.158, and 0.1585 at $b=0.36$;
0.15, 0.151, 0.152, 0.153, 0.154, 0.155, and 0.1555 for $b=0.37$.
The  masses are extracted from the two-point correlation functions 
$C(t)$ by fitting them to a sum of three exponentials. As a matter of 
fact, what we fit is $M_{\mathrm{eff}}(t)=\log(C(t-1)/C(t))$ in the range 
$t\in\{t_{\mathrm{min}},13\}$. For the pion mass we use the
pseudoscalar operator $\Psibar\gamma_5\Psi$ and $t_{\mathrm{min}}=3$. For the 
rho mass we use the vector current $\Psibar\gamma_i\Psi$ and fit from 
$t_{\mathrm{min}}=2$.  In all cases good fits are obtained, with reduced chi-squares
below 1 for the pion and below 1.2 for the rho.

\begin{figure}[!tbp]
\includegraphics[width=\textwidth]{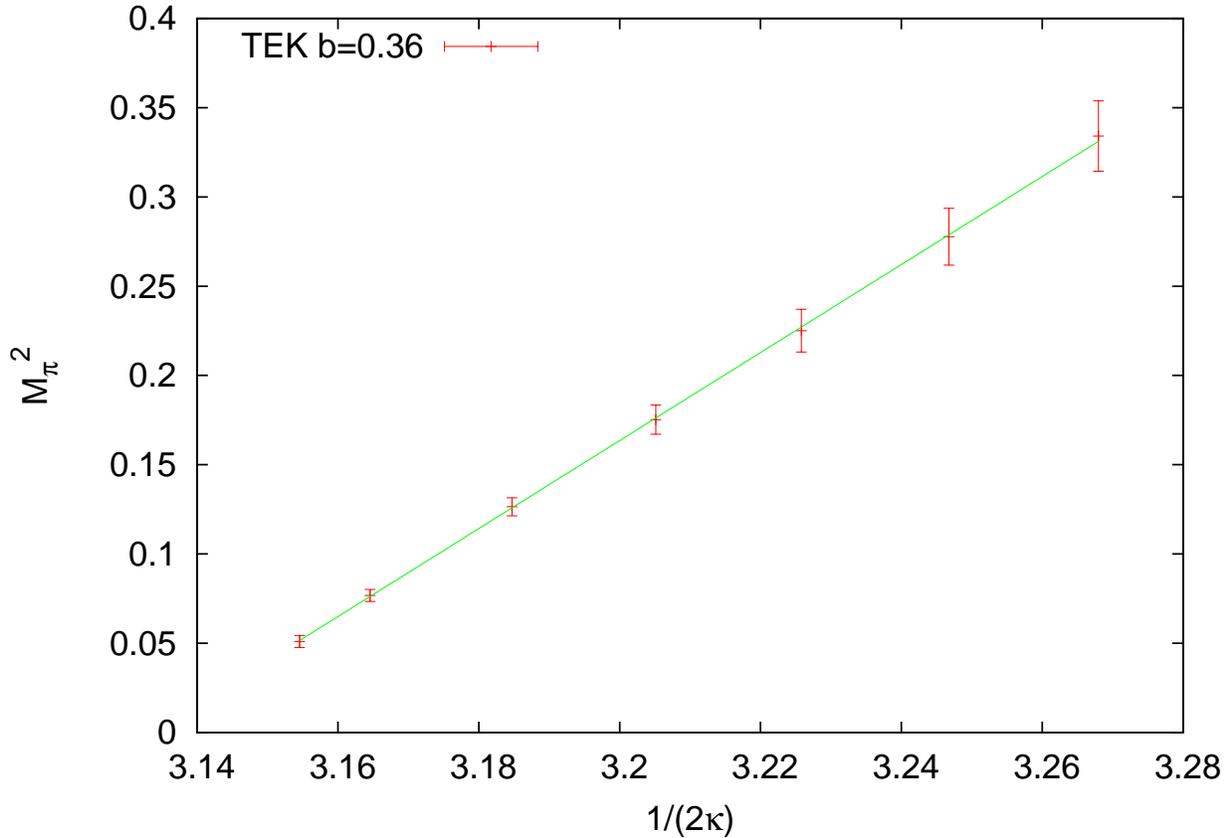}
\caption{The pion mass square in lattice
units is plotted as a function of $1/(2\kappa)$. The continuous line 
is a linear fit.}
\label{fig1}
\end{figure}

The pion mass follows the characteristic chiral symmetry breaking
pattern, with the mass square depending linearly on $\frac{1}{2
\kappa}$. This is shown in Fig.~\ref{fig1} for b=0.36. This allows a
determination of the critical hopping parameter $\kappa_c$ at which the 
pion mass vanishes. In the case of the rho mass a linear fit of the
mass works well and gives a non-zero value for the lattice mass at 
$\kappa=\kappa_c$, which we label $M_{\rho}^{(0)}(b)=m_{\rho}^{(0)}
a(b)$ where $a(b)$ is the lattice spacing. For $b=0.36$ 
we have $\kappa_c=0.1596$ and $M_{\rho}^{(0)}(0.36)=0.391(3)$. For 
$b=0.37$ we have $\kappa_c=0.1559$ and $M_{\rho}^{(0)}(0.37)=0.230(3)$.
The numbers at different $b$ cannot be compared without knowing the
value of the lattice spacing at those $b$. In our case we can use our 
results for the string tension at infinite $N$ from Ref.~\cite{string-tension} to give the 
dimensionless ratio $m_\rho/\sqrt{\sigma}$, which comes out to be 
1.90 and 1.46 for $b=0.36$ and $0.37$ respectively. This implies
considerable scaling violations at these $b$ values. The errors given 
before are only statistical and we did not attempt a thorough estimate
of the systematic errors. 

\begin{figure}[!tbp]
    \begin{minipage}[b]{0.47\textwidth}
\includegraphics[width=\textwidth]{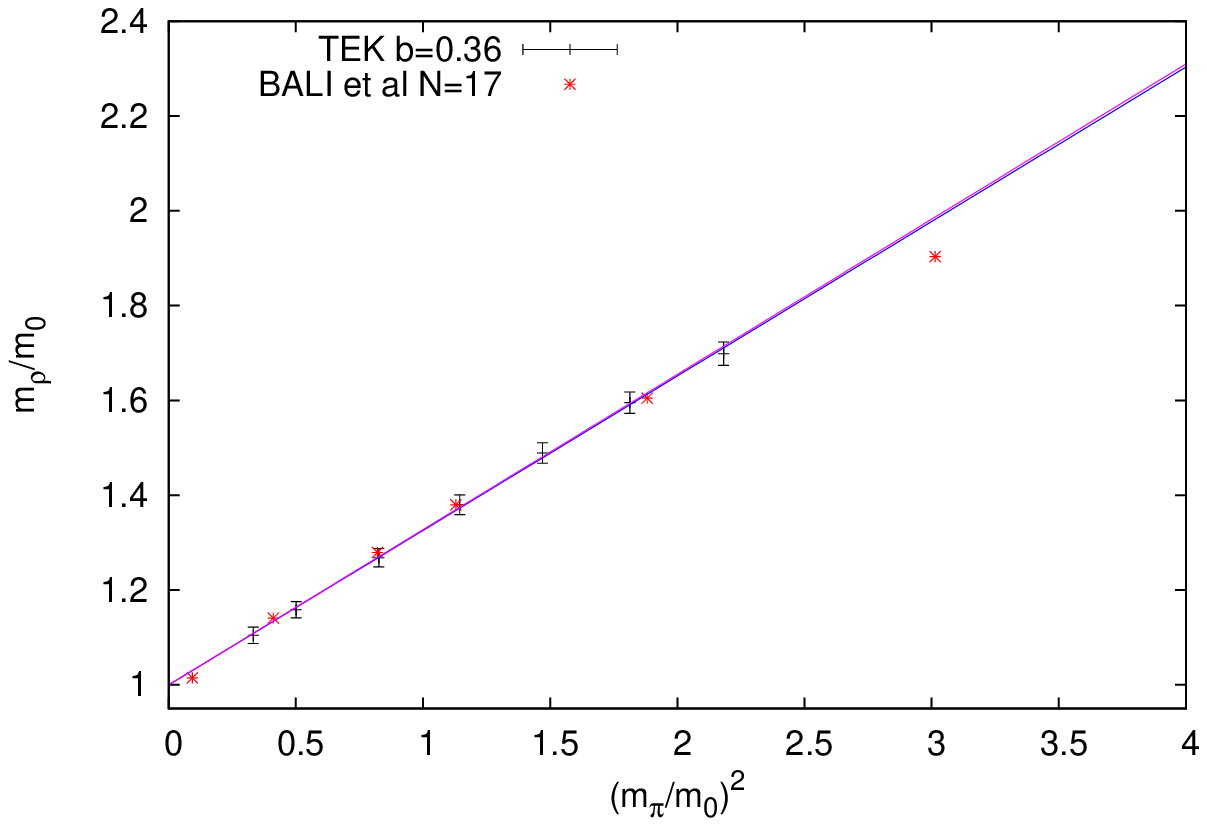}
\caption{The $\rho$ mass versus the square of the pion
mass for our $b=0.36$ data. All masses in units $m_0=m_\rho^{(0)}$. 
The plot also shows the data of Ref.~\cite{lucini} and linear fits to
both sets of data.}
\label{fig2}
\end{minipage}
\hfill
\begin{minipage}[b]{0.47\textwidth}
\includegraphics[width=\textwidth]{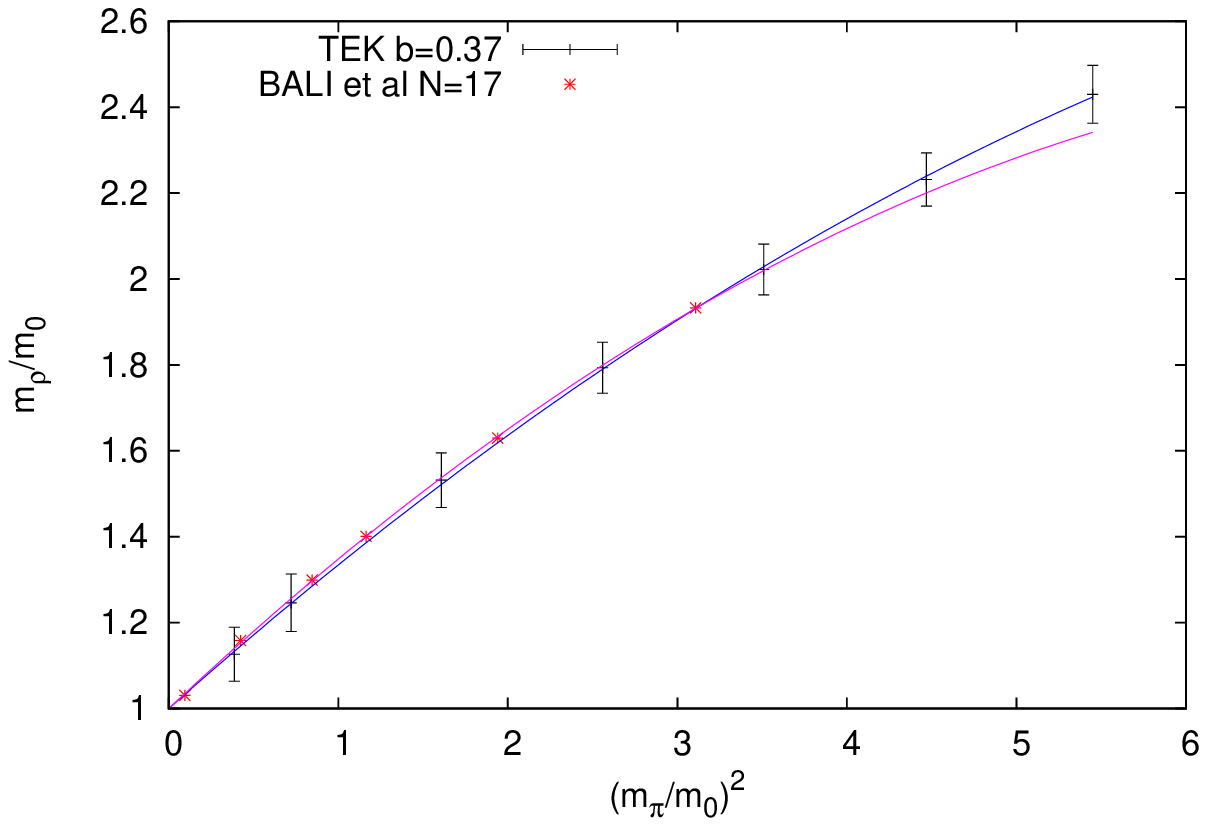}
\caption{The same as Fig.~\ref{fig2} for our b=0.37 data. Fits 
are now given to quadratic functions.\\
\hfil \\
\hfil}
\label{fig3}
\end{minipage}
\end{figure}

It is also interesting to compare our results 
with those of other authors. We chose to take those of Bali et
al~\cite{lucini} corresponding to $N=17$. Their results are obtained at 
$b=0.36064$, which is not far from our value of $0.36$. Nevertheless,
a way to circumvent the difference in scale is to use $m_\rho^{(0)}$
to fix the scale. This might take away part of the systematic error. 
In Figs~\ref{fig2}-\ref{fig3} we display $m_\rho/m_\rho^{(0)}$ as a 
function of $(m_\pi/m_\rho^{(0)})^2$. For small pion masses this
quantity behaves as a straight line with intercept at 1. Our result at
b=0.36 is displayed in Fig.~\ref{fig2} and compared with the same
determination from Ref.~\cite{lucini} for N=17. The result is in perfect
agreement. The fitted slope  for our data is 0.326, and that for the
values of Ref.~\cite{lucini}  is 0.3275. The precision is somewhat
accidental since a quadratic fit covering the full range of the data of
Bali et al gives a slope at the origin of 0.371. Our data for b=0.37, displayed in 
Fig.~\ref{fig3}, also demand a quadratic fit giving a slope at the
origin of 0.351. It is clear from the figure, that errors are much
larger in this case. For the purpose of this letter we are satisfied to
see that our method produces very reasonable numbers with fairly
limited computational resources.

We now summarize our conclusions. 
In this letter we showed how one can compute the meson masses at large
$N$ using a single-site lattice model. The formula is quite simple, and
using it we obtained  results which are very similar to those obtained by
other methods with a very limited amount of resources. In this work we
opted for brevity, focusing on deriving and testing the formula for Wilson
quarks in the fundamental representation in the 4-dimensional SU($N$) theory
at large $N$. The idea and methodology can be readily extended to
different kinds of lattice fermions, arbitrary representations of
SU($N$), and different space-time dimensions. Currently we are applying
the method to `t Hooft model in two dimensions~\cite{GPGAO}, in
which the spectrum is known and the lattice literature is fairly
scarce. Apart from testing the formula it enables to develop the
technology for a precise determination in four dimensions including
scalar and tensor meson masses and decay constants. There is indeed 
no problem in defining smeared meson operators in the reduced model~\cite{okawaproc}.

As mentioned earlier, the case of quarks in other 
representations can be addressed similarly. Despite the fact that 
gauge fields are defined in a one-point lattice, we should allow for 
quarks propagate in a larger lattice. This 
removes the conflict with twisted boundary conditions. Particularly 
interesting are the two index representations (symmetric,
anti-symmetric and adjoint). However, in those cases quarks are
dynamical, and the numerical task of generating thermalized
configurations is much more demanding. The case of quarks in the
adjoint is particularly simple since fermions can be fully reduced to a 
single-site~\cite{GAO} and dynamical quark configurations with two 
flavours are already available~\cite{nf2}.

\acknowledgments{
We acknowledge financial support from the MCINN
grants FPA2012-31686 and FPA2012-31880,
and the Spanish MINECO's ``Centro
de Excelencia Severo Ochoa'' Programme under grant
SEV-2012-0249. M. O. is supported by the Japanese MEXT grant No
26400249 and the MEXT program for promoting the enhancement of research
universities.
Calculations have been done on Hitachi SR16000 supercomputer
both at High Energy Accelerator Research Organization(KEK) and YITP in
Kyoto University. Work at KEK is supported by the Large Scale Simulation
Program No.15/16-04. 
}

\end{document}